\documentclass[aps,pra,twocolumn,showpacs,groupedaddress]{revtex4}  
\usepackage{graphicx}
\usepackage{dcolumn}
\usepackage{bm}
\usepackage{hyperref}
\usepackage{lineno}
\usepackage{color}
\begin{document}
\title{Emergence of extreme events in a quasi-periodic oscillator} 

\author{Premraj Durairaj$^1$, Sathiyadevi Kanagaraj$^1$, Suresh Kumarasamy$^1$, Karthikeyan Rajagopal$^{1,2}$}
\affiliation{$^{1}$Centre for Nonlinear Systems, Chennai Institute of Technology, Chennai-600 069, Tamilnadu, India.\\
$^{2}$Department of Electronics and Communications Engineering, University Centre for\\ Research and Development, Chandigarh University, Mohali, 140 413, Punjab, India.}
\received{:}
\date{\today}
\begin{abstract}
\par Extreme events are unusual and rare large-amplitude fluctuations that occur can unexpectedly in nonlinear dynamical systems. Events above the extreme event threshold of the probability distribution of a nonlinear process characterize extreme events. Different mechanisms for the generation of extreme events and their prediction measures have been reported in the literature. { Based on the properties of extreme events, such as rare in the frequency of occurrence and extreme in amplitude, various studies have shown that extreme events are both linear and nonlinear in nature.} Interestingly, in this work, we report on a special class of extreme events which are nonchaotic and nonperiodic. These nonchaotic extreme events appear in between the quasi-periodic and chaotic dynamics of the system. We report the existence of such extreme events with various statistical measures and characterization techniques. 
\end{abstract}
\maketitle 
\par Extreme events are unanticipated, rare events that occur in many natural and engineering systems. Extreme events (EE) can exist in various forms, including floods, cyclones, droughts, pandemics, power outages, material ruptures, explosions, chemical contamination, and stock market crashes, among others \cite{albeverio06}. Such events have a severe impact on real-world situations. Thus, it is necessary to understand the relevant mechanism and its generic characteristics for the occurrence of EE in order to prevent such EE. As a result, the researchers focused on exploring the EE in diverse nonlinear oscillators \cite{hohmann10,metzger14,mathis15,birkholz16}, maps \cite{prem1}, and neural networks \cite{saha}.  Further, the extreme events have also been identified in a super-fluid helium \cite{ganshin08}, plasma\cite{bailung11}, optical fibers \cite{solli07}, lasers \cite{zamora13}, and capillary wave \cite{Onorato} etc. 

\par However, depending on the characteristics of a dynamical system, the occurrence of EE has been discovered under a variety of mechanisms, including internal crises, on-off intermittency, blowout bifurcations, stick-slip bifurcations, and so on \cite{prem1, Venk, Grebogi, zamora13, Reinoso}. For instance,  prior studies reveal that EE can arise as a result of the abrupt expansion and destruction of chaotic attractors produced by internal or external crises \cite{zamora13, Grebogi}. Further, interior crises are found to be a critical mechanism for the occurrence of EE, when the trajectory of chaotic attractors reaches the stable manifold of a saddle or unstable periodic orbit, which increases the size of the chaotic attractors. Such a sudden expansion of the chaotic attractor may result in EE. In addition, Pomeau-Manneville intermittency is identified as another mechanism for the existence of EE. Such intermittency can occur when the periodic oscillations are interspersed by chaotic bursts, which further results in very large amplitude events. EEs can also exist through the following other mechanisms. The sliding bifurcation near the discontinuous boundary can cause EE. The trajectory of the attractors might hop between coexisting attractors due to noise in multi-stable systems, which can cause unusually large events. This is referred to as noise-induced intermittency. The trajectory of the attractors in coupled systems departs from the synchronization manifold to the transverse direction of the manifold. During such a transition, a synchronization error of dynamics can show zero or nonzero and is referred to as on-off intermittency \cite{review}. 

\par Moreover, previous studies discovered that extreme or rare events can occur as a result of chaotic or stochastic processes \cite{review}. In particular, the appearance of EE has been reported in micro-electromechanical cantilevers with discontinuous boundaries and diode lasers with phase-conjugate feedback \cite{Mercier,Kumarasamy}.  By applying the harmonic pump modulation to the fiber laser the emergence of Rogue waves has been identified \cite{Pisarchik1,Pisarchik2,rw2,hohmann10}. The EE in stochastic transport on networks has been demonstrated using multiple random walks on complex networks \cite{Santhanam1,Santhanam2}. Now the interesting question is whether extreme events can be induced by nonchaotic signals. In literature, a study has shown nonchaotic and nonperiodic have been well studied in the name of strange nonchaotic dynamics, which arises during the attractor transition from quasi-periodicity to chaos \cite{Pikovsky}. One can find the generation mechanisms of these strange nonchaotic attractors in literature \cite{Pikovsky,AP,prem2}. The results in the present work show that similar to the strange nonchaotic dynamics, the nonperiodic and nonchaotic dynamics show large-amplitude extreme events. The present study opens a new area of study where the nonchaotic nonlinear process can also lead to extreme events and the same has not been found reported. 

\par To show the nonchaotic extreme events, we consider the Morse Oscillator (MO) which is used to describe the motion of diatomic molecules. Importantly, the MO has made substantial contributions in the fields of classical, semi-classical, and quantum mechanics \cite{Lie,Knop}. The MO was used for photo-dissociation molecules without any damping.  In the presence of driving and damping, the MO was exploited for multi-photon excitation of molecules, pumping the local mode of polyatomic molecules \cite{Zdravkovic}. 
We consider the quasi-periodically forced MO and its dynamical equation can be written as 
\begin{eqnarray}
	\dot{x} &=& y \nonumber\\
	\dot{y}&=& fsin(\omega_{1}t)+gsin(\omega_{2}t)+ e^{-2x}-e^{-x}-\gamma y
	\label{eq1}
\end{eqnarray}
where $x$, and $y$ are the state variables of the system and $\gamma$ is a damping parameter. The amplitudes of the first and second force are represented by $f$ and $g$ and the corresponding frequencies are denoted by $\omega_1$ and $\omega_2$, respectively. 
\begin{figure}[!h]
\begin{center}
\includegraphics[width=0.48\textwidth]{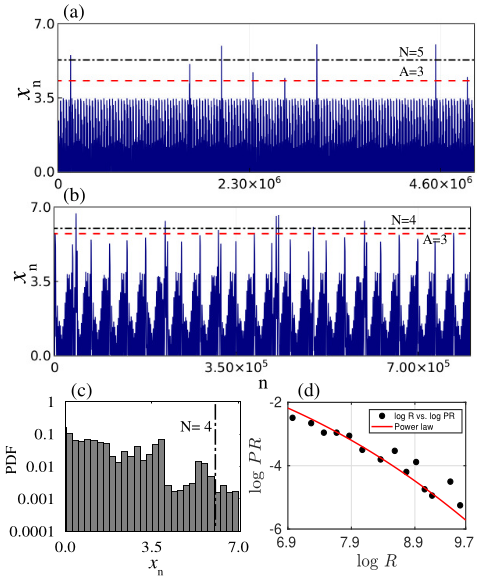}
\end{center}
\caption{{{(a) Time evolution of $x_{n}$ for the nonchaotic dynamics with forcing amplitudes as (a) $f=g=0.255$, and (b)  $f=g=0.278$. The $x_n$ is the $n^{th}$ local peaks of the variable $x$. The horizontal black dot-dashed and red dashed lines are the critical threshold lines defining the extreme events (refer to the text for the meaning of $N$ and $A$).}} (c) The probability distribution function corresponds to the extreme events and (d) return interval (R) (inter-event interval) with respect to the probability of recurrence times (PR) of the EE for (b). The filled circles and solid lines in (d) represent the numerical data and the corresponding power-law fit. We fixed the other parameter values as $\gamma=0.35$, $\omega_{1}=0.3$, and $\omega_{2}=(\frac{\sqrt 5-1}{2}$). } 
\label{ts}
\end{figure}
\par { {To manifest the existence of extreme events, we first depicted the time evolution of the $x$-variable in Fig.~\ref{ts}(a) and  Fig.~\ref{ts}(b) by fixing the amplitude of the first and second forcing as $f=g=0.255$ and  $f=g=0.278$. We observe from Fig.~\ref{ts}(a) that some of the oscillation(event) has larger amplitudes, while the rest of them take lower amplitudes. }To check the larger amplitude oscillations satisfy the extreme events criteria defined in the literature, we use the following relation:
\begin{eqnarray}
x_{_{EE}}=<x_n> + N \sigma_{x{_n}},
\label{ct}
\end{eqnarray}
where $x_{_{EE}}$ is the critical amplitude threshold and $N$ is a multiplication factor. The mean and standard deviation of the variable $x$ is represented by $<x_n>$ and $\sigma_{x{_n}}$, respectively. Here, the $x_n$ (an event) are the local peaks of the variable $x$. An event or a local peak can satisfy extreme event criteria if it has a value higher than the critical threshold defined by Eq.~(\ref{ct}) with $N\ge4$. {To confirm the presence of EE, we plotted the critical threshold on the time series for $N=5$ and $N=4$ in Figs. \ref{ts}(a) and \ref{ts}(b). We used two different $N$ values depending on the time series. Though the choice of $N$ is arbitrary, we set the minimum $N$ value as 4 in the present study. We also find the critical value of $N_{max}$ for a range of each $f$ value -- the details will be discussed below.  In both cases, we can see that some of the large amplitude events cross the threshold line, confirming the presence of EE. Since the choice of $N$ is arbitrary in the previous criterion, we use another criterion defined by the abnormality index; $A_n=\frac{Hf_n}{H_{1/3}}$ \cite{Mercier}, where $Hf_n$ is the difference between the maximum height of the event $n$ and the mean height of its population,  $Hf_n=x_n-\langle x_n\rangle_n$ and $H_{1/3}$ is the average value among the highest one-third values of $Hf_n$. If an event $x_n$ has abnormality $A_n$ greater than  2 then the event is termed an extreme event. We find that both cases in Figs. \ref{ts}(a) and \ref{ts}(b) satisfy the above criterion with abnormality index $A=3$ denoted by a dashed horizontal line in the plots.  It is evident that a few rare large amplitude events cross the abnormality index line.} 
We computed the probability distribution function (PDF)  in Fig.~\ref{ts}(c)  for the time series shown in Fig.~\ref{ts}(b).  The EE critical threshold at $N=4$ is plotted as a vertical dashed line on the PDF diagram. In the plot, the events with a finite probability above the critical threshold line characterize the extreme events. We can plot similar probability distribution for Fig. 1(a), however, for simplicity, we have plotted the PDF corresponding to Fig. 1(b).}

\par {The above analysis shows that the observed behavior satisfies the extreme events criterion in the amplitudes. Another important characteristic of extreme events is an inter-event interval. The inter-event interval defines the frequency occurrence of the events and should not have discrete values (discrete values mean the periodic occurrence of events), rather it should have a distribution over a range. In order to examine the distribution of events in the observed time series, we find inter-event intervals (R) between successive extreme events. Subsequently, we find the probability of such inter-event intervals (PR) as shown in Fig.~\ref{ts}(d). Inter-event interval and its probability obey power-law relations as given by $log_{10}(PR) = a ~log_{10} (R)^{b}$, where $a$ and $b$ are constants with values $a=-0.006$ and $b=2.96$, respectively. The obtained numerical values are depicted in a filled circle, and a continuous line shows the corresponding power-law fit. The route for the emergence of EE and its transitions is further estimated below using Lyapunov exponents (LE), amplitude maxima $X_{max}$, critical factor $N_{max}$, and two-parameter analysis. } 

\begin{figure}[ht!]
\begin{center}
\includegraphics[width=0.45\textwidth]{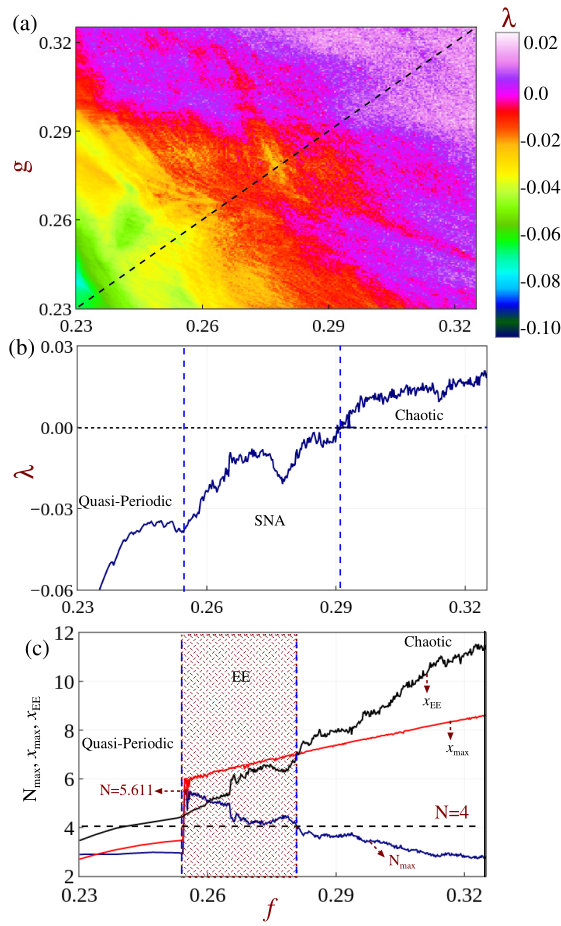}
\end{center}
\caption{ (a) The two-parameter bifurcation diagram in $(f,g)$ space.  Using the range of Lyapunov exponents ($\lambda$) (denoted by the color bar) the dynamical regions are marked. (b) The maximum Lyapunov exponents as a function of forcing amplitude $f(=g)$, { (c) maximum amplitude of the events $x_{max}$ (red) and the corresponding  $N_{max}$ (Eq. \ref{nmax}) of the event (blue) by varying the magnitude of $f(=g)$.  The black line represents the extreme events critical threshold ($x_{EE}$) drawn from Eq.~(\ref{ct}) for N=4.  The other parameter values are fixed as the same as in Fig.~\ref{ts}.} }
\label{le}
\end{figure}
\par {To illustrate the global dynamical transition of the attractors and route of the EE, the two-parameter diagram is drawn in $(f,g)$ space using the maximum LE as shown in Fig.~\ref{le}(a). The range of LE (shown in the color bar) denotes the emergence of quasi-periodic, nonchaotic, and chaotic attractors in the respective parameters of $f$ and $g$.  If the forcing amplitudes $f$ and $g$ are small, attractors have a maximum negative LE, indicating the presence of a quasi-periodic (QP) attractor region. To better comprehend QP attractors, we plotted their time-evolution and phase portrait trajectories in Supplementary Material Fig.~S1 a(i, ii) for $f=g=0.23$, which show their bounded nature. Thus, the EE critical threshold for this attractor is greater than the amplitude of QP attractors. By increasing $f$ and $g$ values, the QP attractor transits to a chaotic (CH) attractor via strange and nonchaotic dynamics in which the LE takes the values from negative (near zero) to positive. {To distinguish between the strange nonchaotic and chaotic attractors, the time-evolution and phase portrait trajectories are shown in Figs.~S1 b(i,ii) and Figs.~S1 c(i,ii) in the supplementary materials by fixing $f=g=0.278$ and $f=g=0.33$, respectively. { Also, the frequency spectra can be used to distinguish quasiperiodic, SNA and chaos. We have the frequency spectrum analysis in the Supplementary material in Fig.~S4 (a-c)}.  When compared to the chaotic attractor (which has a greater number of large amplitude oscillations), we found the SNA shows fewer large amplitude oscillations. The supplemental material's Fig. S1 can be consulted for more information.} Furthermore, to show the dynamical transitions clearly, we displayed maximum Lyapunov exponents in Fig.~\ref{le}(b) by keeping the parameter ($f=g$) and varying it along the diagonal dashed line shown in Fig.~\ref{le}(a). In  Fig.~\ref{le}(b), the maximum LE is illustrated as a function of forcing amplitudes $f$ and $g$ ($f=g$) in the range $(0.23 <f(=g)< 0.32)$. We observe that when the forcing amplitudes are minimum in the mentioned range, LE takes negative values, indicating quasi-periodic dynamics. While increasing the parameter, the transition of LE from negative to positive values indicates the dynamical transition of quasi-periodic behavior to chaotic behavior. Furthermore, we found that the negative values of LE near-zero exhibit strange nonchaotic behavior; extreme events are seen in this region. {The literature has shown that the EEs occur under chaotic dynamics \cite{review} through distinct routes and stochastic processes like stochastic transport on networks has been demonstrated using multiple random walks on complex networks \cite{Santhanam1,Santhanam2}}. Among the various routes, the occurrence of EEs in nonchaotic dynamics is new and it has not been reported to the best of our knowledge.}

{ To validate the occurrence of EEs in the SNA region, we find the maximum amplitude $x_{max}$, extreme event threshold $x_{EE}$, and maximum value of $N$ ($N_{max}$) of a given time series. In Fig.~\ref{le}(c), we have plotted the above quantities by varying the magnitude of $f=g$. The plot explains the regime of extreme events in the following way. During the non-extreme regime, the critical threshold $x_{EE}$ is larger than the $x_{max}$. It means that the threshold is larger than the large amplitude oscillations and does not satisfy the extreme events criterion. While in the EE regime, the $x_{max}$ is larger than the EE critical threshold $x_{EE}$ (shaded EE region). This explains that extreme events have a larger amplitude than the extreme event criterion. Note that the SNA regime in the parameter range $f \in{0.28}$~ to~${0.2912}$ shows no extreme events.  As we discussed above, we fixed $N = 4$ as an arbitrary constant from the literature \cite{Bonatto}. However, the maximum value of the $N$ can be determined by rewriting Eq. (2), as }
\begin{eqnarray} \label{nmax}
 N_{max} = \frac{max(x_{EE})-\langle x_n \rangle}{\sigma_{x_n}}.
\end{eqnarray}
\begin{figure}[!h]
\begin{center}
\includegraphics[width=0.48\textwidth]{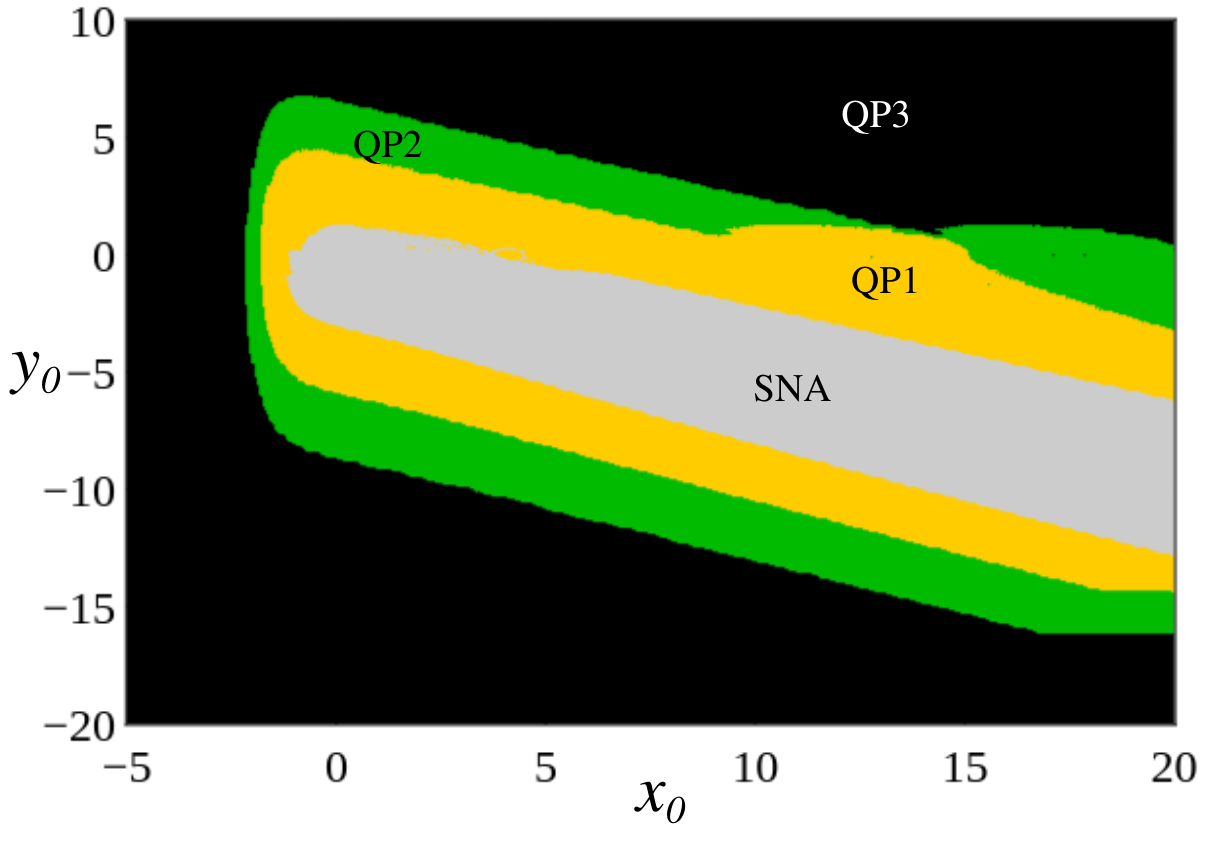}
\end{center}
\caption{Basin of attraction for $f=g=0.278$. $QP1$, $QP2$, and $QP3$ are the quasi-periodic attractor-1, quasi-periodic attractor-2, and quasi-periodic attractor-3, respectively.  SNA represents the strange nonchaotic attractor. We fixed the other parameter values the same as in Fig.~\ref{ts}. }
\label{fig:21}
\end{figure}
{ { In the SNA region shown in Fig.~\ref{le}(c),  we found that the multiplication factor taking values between $4 \le N_{max} \le 5.611$ when the forcing amplitudes in the range from $0.256$ to $0.28$ denoted by shaded transparent pattern. {The plot of $N_{max}$ shows that depending on the parameter choice, the arbitrary value can be chosen N$\in\{4,5.611\}$. Thus above results satisfy all the criteria proposed for the extreme events and justify the existence of EEs in the SNA regime.}}}

\par  As we discussed earlier, the observed EEs are nonchaotic and nonperiodic. At the same time, the parameters corresponding to the strange nonchaotic EEs show multiple stable behaviors. The multi-stable behavior can be seen from the basins of attraction drawn for a range of initial conditions. Figure~\ref{fig:21} is drawn by varying the initial states $x_0$ and $y_0$ of the system for the parameters given in Fig.~\ref{ts} caption. We can see that basin of nonchaotic and nonperiodic behavior or SNA is embedded within the basin of quasi-periodic dynamics. Outside the SNA basin, we have found three different basins which contain quasi-periodic attractors. All the three different quasi-periodic attractor basins and the SNA basin, denoted by QP1, QP2, QP3, and SNA respectively in Fig. \ref{fig:21}. In supplemental material Fig. (S2), each of the quasi-periodic attractors is depicted. Figure \ref{fig:21} shows that extreme events occur for specific values of initial conditions. The size of these basins changes as we vary the parameter within the EEs regime marked in Fig.~\ref{le}. 
\begin{figure}[ht]
\begin{center}
\includegraphics[width=0.45\textwidth]{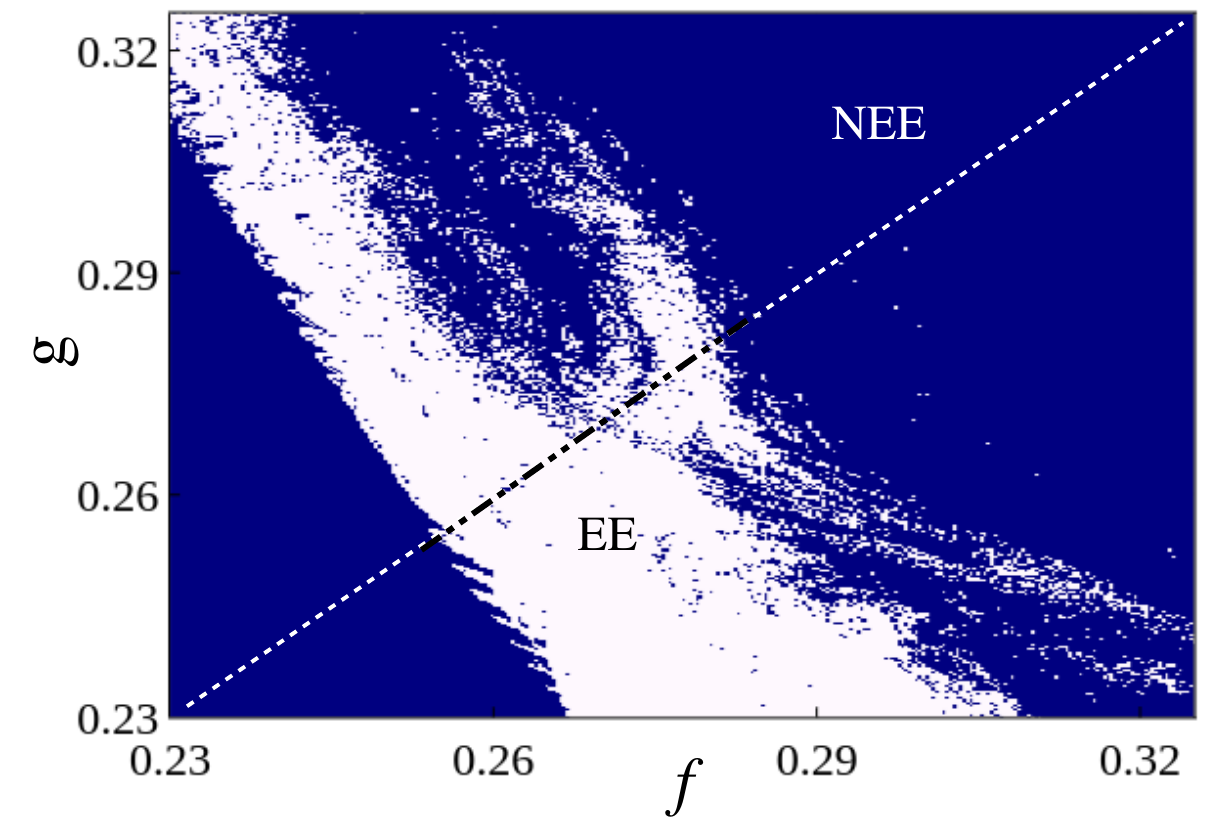}
\end{center}
\caption{Two parameter phase diagram in $(f,g)$ space (plotted using Eq.~\ref{ct} for fixed initial condition $(x_0,y_0)=(0.3,0.2)$), to distinguish the existence of extreme events (EE) and non-extreme events (NEE), respectively.  We fixed the other parameter values as the same as in Fig.~\ref{ts}.}
\label{fig:3}
\end{figure}

\par Similarly, to determine the regime of the extreme event in the parametric space between $f$ and $g$, a two-parameter diagram is drawn as shown in Fig.~\ref{fig:3}. The white regime in the plot shows the extreme events for the combinations of parameter $(f,g)$ separated with the help of Eq. ~\ref{ct} from the non-extreme events (NEE-- denoted by blue color). By comparing Fig.~\ref{le}(a) with Fig.~\ref{fig:3} we can say that EEs occur in the SNA region (however some of the SNA parameter regime may not contain EEs).

\begin{figure}[ht!]
	\begin{center}
		\includegraphics[width=0.5\textwidth]{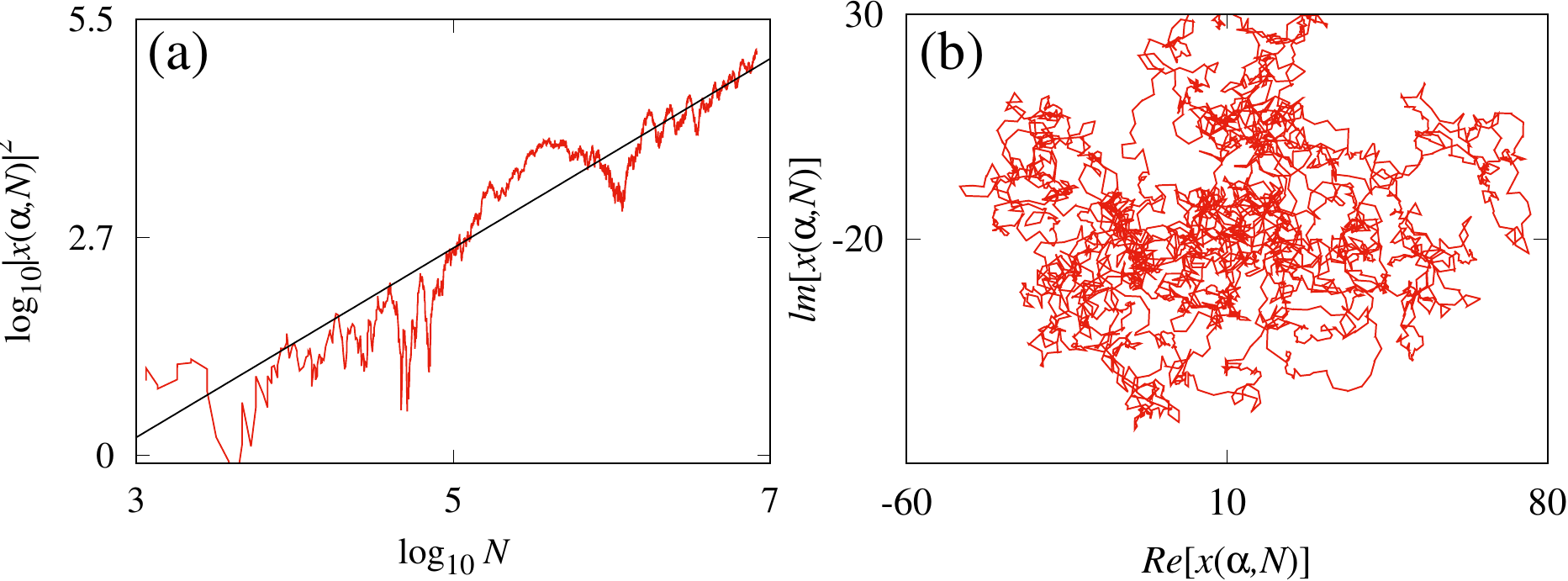}
	\end{center}
	\caption{Singular continuous spectrum for fixing the forcing amplitudes $f,g$ $=$ $0.278$. (a) The logarithmic plot of $|x(\alpha,N)|^2$ against $N$. The red and black lines denote the numerical values and the corresponding power-law fit. (b) Fractal path in the complex plane of $x$. The other parameter values are defined as $\gamma=0.35$, $\omega_{1}=0.3$, $\omega_{2}=(\frac{\sqrt 5-1}{2}$).}
	\label{fig:5}
\end{figure}

To show the generality of the existence of EEs in the SNA regime, we present the regime of EEs for $\gamma=0.4$ in the supplementary material Figs.~S3 (a),(b). This result validates the presence of strange nonchaotic extreme events in the selected parameter regime. In the following section, we characterize the observed behavior as strange and nonchaotic in nature. For this purpose, we perform singular continuous spectrum analysis and distribution of finite-time Lyapunov exponents. 	



\par To validate the strange nonchaotic dynamics, we plot singular continuous spectrum \cite{Pikovsky} in Fig.~\ref{fig:5} using partial Fourier sum of the signal $x$ given by $X(\alpha,N)=\sum_{m=1}^{N}x_m e^{2\pi i m \alpha}$, where $\alpha$ is proportional to the external frequency ($\omega_{1}$) and $N$ is the length of the time series. The red and black lines show the singular continuous spectrum and the corresponding power-law fit. When N is considered as time, $|X(\alpha,N)|^2$ grows with
N, that is $|X(\alpha,N)|^2\sim N^{\beta}$, where $\beta$ is the slope. When the signal possesses the properties of strange nonchaotic dynamics, the corresponding slope values lie between $1<\beta<2$. For this case, the slope value $\beta=1.576$ confirms the existence of strange nonchaotic dynamics shown in Fig. \ref{fig:5}(a). The corresponding path of Brownian motion with fractal structure in complex [$Re(x), Im(x)$] plane also confirms the strange nonchaotic dynamics in Fig. \ref{fig:5}(b).    

\begin{figure}[!ht]
	\begin{center}
		\includegraphics[width=0.5\textwidth]{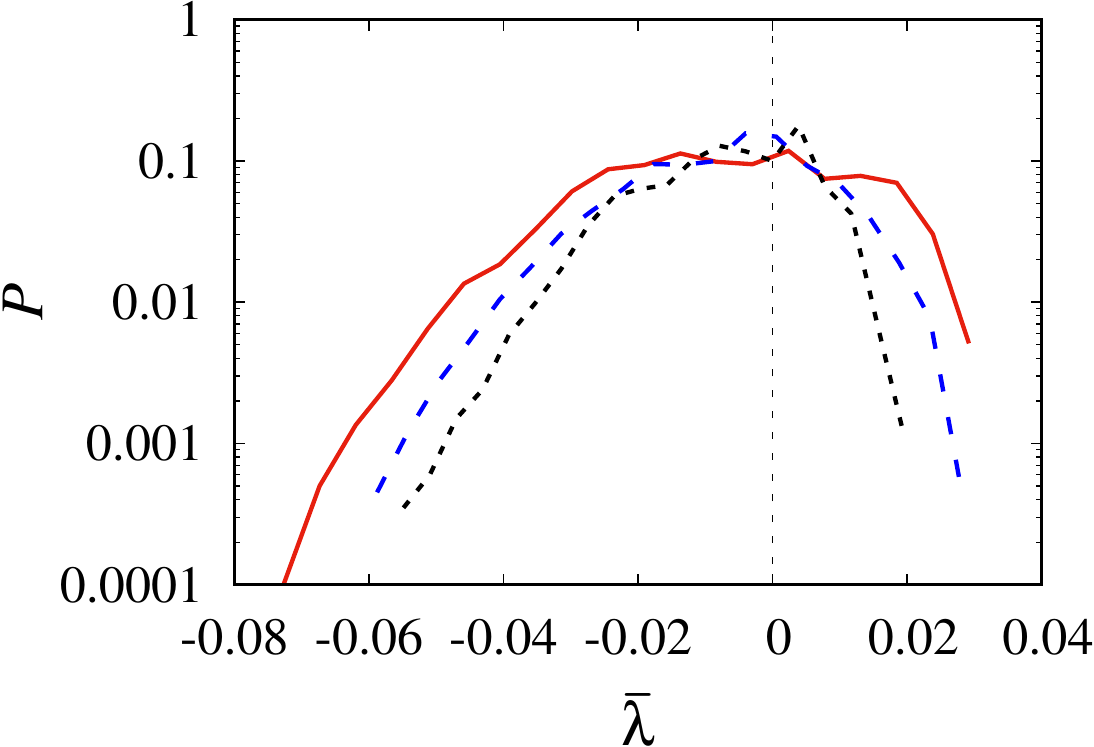}
	\end{center}
	\caption{Finite time Lyapunov exponent with respect to probability distribution function (PDF) for SNAs by fixing the three distinct finite time periods $T= 500$ (red line), $T = 1000$ (blue dashed line), and $T = 1500$ (black dotted line) with $f,g = 0.278$.}
	\label{fig:6}
\end{figure}

\par The strange nonchaotic dynamics are also validated using another statistical characterization known as the distribution of finite-time Lyapunov exponents. The distribution takes both positive and negative values, but the area under the curve is maximum in the negative regime for strange nonchaotic dynamics. Figure~\ref{fig:6} plotted for three different finite time intervals $T=500,~1000$, and $1500$, the distribution has a large negative region compared to the positive region showing nonchaotic dynamics. From these analyses, the observed dynamics are strange (nonperiodic) as well as nonchaotic, which also shows the large amplitude and rare events.

\par The present letter shows a mechanism of the emergence of extreme events in a quasi-periodically forced Morse oscillator. As a function of forcing amplitude, we found the transition from quasi-periodic (QP) to chaotic (CH) attractor via strange nonchaotic extreme events. During such extreme event dynamics, we found a long excursion of trajectories that are away from the bounded attractor, while the chaotic attractors show many higher amplitude peaks. To confirm the existence of EEs, we estimated the critical threshold, and it is observed that the higher amplitude peaks in the EE cross the critical threshold while the peaks in the CH and QP attractor do not. The dynamical transitions of the attractors and the occurrence of nonchaotic EE dynamics are manifested through maximum Lyapunov exponents. The observed extreme events are further validated using the probability distribution and return interval (inter-event interval) with respect to the probability of recurrence times of the EE. Extreme events are abnormal and unexpected events that occur in many natural and man-made systems. Understanding the mechanism or route can help to anticipate the onset of EEs. Early works on extreme events show the chaotic nature of the extreme events because of the rare and extreme amplitude properties of extreme events. The present study shows an unknown emergence of extreme events that are nonchaotic and nonperiodic extreme events. This finding shed light on the new direction where extreme events can happen as a nonchaotic process.  
\par We gratefully acknowledge this work is funded by the Center for Nonlinear Systems, Chennai Institute of Technology (CIT), India, vide funding number CIT/CNS/2022/RP-016.


\begin{thebibliography}{100} 
\bibitem{albeverio06}

	S. Albeverio, V. Jentsch and H. Kantz, Extreme Events in Nature and Society, The Frontiers Collection (Springer, Berlin, 2006).

\bibitem{hohmann10} 

	R. H\"{o}hmann, U. Kuhl, H. J. St\"{o}ckmann, L. Kaplan and E. J. Heller, Phys. Rev. Lett., \textbf{104}, (2010) 093901.

\bibitem{metzger14} 

	J. J. Metzger, R. Fleischmann and T. Geisel, Phys. Rev. Lett., \textbf{112}, (2014) 203903.

\bibitem{mathis15}

	A. Mathis, L. Froehly, S. Toenger, F. Dias, G. Genty and J. M. Dudley, Sci. Rep., \textbf{5}, (2015) 12822.

\bibitem{birkholz16} 

	S. Birkholz, C. Br\'{e}e, I. Veseli\'{c}, A. Demircan and G. Steinmeyer, Sci. Rep., \textbf{6},  (2016) 35207.

\bibitem{prem1}

	D. Premraj, K. Suresh, S. A. Pawar, L. Kabiraj, A. Prasad, and R. I. Sujith, EPL, \textbf{134}, (2021) 34006.

\bibitem{saha}

	A. Saha,  and U. Feudel, Phys. Rev. E, \textbf{5}, (2017) 062219.

\bibitem{ganshin08} 

	A. N. Ganshin,  V. B. Efimov, G. V. Kolmakov, L. P. Mezhov-Deglin and  P. V. E. McClintock, Phys. Rev. Lett., \textbf{101}, (2008) 065303.

\bibitem{bailung11}

	H. Bailung, S. K. Sharma and Y. Nakamura, Phys. Rev. Lett., \textbf{107}, (2011) 255005.

\bibitem{solli07}

	D. R. Solli, C. Ropers, P. Koonath and  B. Jalali, Nature, \textbf{450}, (2007) 1054.

\bibitem{zamora13} 

	J. Zamora-Munt, B. Garbin, S. Barland, M. Giudici, J. R. Rios Leite, C. Masoller and J. R. Tredicce, Phys. Rev. A, \textbf{87}, (2013) 035802.

\bibitem{Onorato}

	M. Onorato, T. Waseda, A. Toffoli, L. Cavaleri, O. Gramstad, P. A. E. M. Janssen, T. Kinoshita, J. Monbaliu, N. Mori, A. R. Osborne and M. Serio, Phys. Rev. Lett. \textbf{102},  (2009) 114502.

\bibitem{Venk}

	S. C. Venkataramani, B. R. Hunt, E. Ott, D. J. Gauthier and J. C. Bienfang, Phys. Rev. Lett., \textbf{77}, (1996) 5361.

 
\bibitem{Grebogi}

    C. Grebogi, E. Ott, F. Romeiras and J. A. Yorke,  Phys. Rev. A, \textbf{36},  (1987) 5365.

\bibitem{Reinoso}

	J. A. Reinoso, J.  Zamora-Munt,  and C. Masoller, Phys. Rev. E, \textbf{87},  (2013) 062913.

\bibitem{review}

	S. N. Chowdhury, A. Ray, S. K. Dana and D. Ghosh, Phys. Rep., \textbf{966}, (2022) 1-52.

\bibitem{Mercier}

	E. Mercier, A. Even, E. Mirisola, D. Wolfersberger, and M. Sciamanna, Phys. Rev. E, {\bf 91}, 042914 (2015).

\bibitem{Kumarasamy}

	K. Suresh,  and A. N. Pisarchik, Phys. Rev. E, \textbf{98}, (2018) 032203.

\bibitem{Pisarchik1}

	A. N. Pisarchik, R. Jaimes-Reátegui, R. Sevilla-Escoboza, G. Huerta-Cuellar, and M. Taki, Phys. Rev. Let., {\bf 107}, 274101 (2011).

\bibitem{Pisarchik2}

	A.N. Pisarchik, A.E. Hramov, Multistability in Physical and Living Systems: Characterization and Applications (Springer, 2022).


\bibitem{rw2}
 F. T. Arecchi, U. Bortolozzo, A. Montina, and S. Residori, Phys. Rev. Lett., {\bf 106}, 153901 (2011).

 
\bibitem{Pikovsky}

	A. S. Pikovsky, U. Feudel and S. P. Kuznetsov, Strange nonchaotic attractors: Dynamics between order and chaos in quasi periodically forced systems (World Scientific (2006)).

\bibitem{Santhanam1}

	V. Kishore, M. S. Santhanam, and R. E. Amritkar, Phys. Rev. let., {\bf 106}, 188701 (2011).

\bibitem{Santhanam2}

	A. Kumar, S. Kulkarni and M. S. Santhanam, Chaos, {\bf 30}, 043111, (2020).

\bibitem{AP}

	A. Prasad, S. S. Negi and R. Ramaswamy, Inter. J. Bifur. and Chaos, \textbf{11}, (2001) 291-309.

\bibitem{prem2}

	D. Premraj, K. Suresh, J. Palanivel and K. Thamilmaran, Communica.  Non. Scie.  Numer. Simula., \textbf{50}, (2017) 103.; D. Premraj, K. Suresh, K. Thamilmaran and K. Rajagopal, The European Phys. J. Spe. Top.,  (2022) 1-7.

\bibitem{Lie}

	G. C. Lie and J. M. Yuan, The J.  chem. phys., \textbf{84}, (1986) 5486.

\bibitem{Knop}

	W. Knop and W. Lauterborn, The J. chem. phys., \textbf{93}, (1990) 3950.

\bibitem{Zdravkovic}

	S. Zdravkovic, A. N. Bugay and A. Y. Parkhomenko,  Nonlinear Dyn., \textbf{90}, (2017) 2841.

\bibitem{Bonatto}

	C. Bonatto and A. Endler, Phys. Rev. E, \textbf{96}, (2017) 012216.


\bibitem{Pikovsky}
A. S. Pikovsky and U. Feudel, J. Phys. A, \textbf{27}, (1994) 5209.


\end{thebibliography}
\end{document}